\documentclass[preprintnumbers,amsmath,11pt,amssymb,floatfix,superscriptaddress,nofootinbib]{revtex4}
\usepackage{graphicx}
\usepackage{epsfig}
\usepackage{bm}
\usepackage{amsfonts}

\input epsf

\begin{document}

\title{\Large A study on the interacting Ricci dark energy in $f(R,T)$ gravity}

\author{Surajit Chattopadhyay}
\email{surajit_2008@yahoo.co.in, surajcha@iucaa.ernet.in}
\affiliation{Pailan College of Management and Technology, Bengal
Pailan Park, Kolkata-700 104, India.} \footnote{The author is a
Visiting Associate of the Inter-University Centre for Astronomy
and Astrophysics (IUCAA), Pune, India.}

\date{\today}

\begin{abstract}
The present work reports study on the interacting Ricci dark
energy in a modified gravity theory named $f(R,T)$ gravity. The
specific model $f(R,T)=\mu R+\nu T$ (proposed by R. Myrzakulov,
arXiv:1205.5266v2 \cite{Myrza1}) is considered here. For this
model we have observed a quintom-like behavior of the equation of
state (EoS) parameter and a transition from matter dominated to
dark energy density has been observed through fraction density
evolution. The statefinder parameters reveal that the model
interpolates between dust and $\Lambda$CDM phases of the universe.
\end{abstract}


\maketitle

\section{Introduction}

The origin of dark energy (for review see \cite{Copeland,
Paddy,BambaOdintsov}) responsible for the cosmic acceleration
\cite{Spergel, Perlmu} is one of the most serious problems in
modern cosmology. The first step toward understanding the nature
of dark energy is to clarify whether it is a simple cosmological
constant or it originates from other sources that dynamically
change in time \cite{Tsuza}. In an extensive review, Nojiri and
Odintsov \cite{Nojiri1} thoroughly discussed the reasons why
modified gravity approach is extremely attractive in the
applications for late accelerating universe and dark energy. Other
remarkable reviews on modified gravity are \cite{Clifton, Tsuza}.
Various modified gravity theories have been proposed so far. These
include, $f(R)$ \cite{Nojiri2, Nojiri3}, $f(T)$ \cite{Cai1,
Ferraro, Bamba1,BambaGeng}, $f(G)$ \cite{Myrza2, Setare1},
Horava-Lifshitz \cite{Kiritsis, Nishioka} and Gauss-Bonnet
\cite{Nojiri4, Li} theories. The present work concentrates on
$f(R,T)$ gravity, with $T$ being the trace of stress-energy
tensor, manifesting a coupling between matter and geometry. Before
going into the details of $f(R,T)$ gravity,
 let us first briefly survey the $f(R)$ gravity. The recent motivation for studying $f(R)$ gravity
has come from the necessity to explain the apparent late-time
accelerating expansion of the Universe \cite{Clifton}. Some
extensive reviews of $f(R)$ gravity are \cite{Soti, Antonio,
Soti1, Capoz}. Thermodynamic aspects of $f(R)$ gravity have been
investigated in the works of \cite{Bambathermo, Akbar}. A
generalization of $f(R)$ modified theories of gravity including in
the theory an explicit coupling of an arbitrary function of the
Ricci scalar $R$ with the matter Lagrangian density
$\mathcal{L}_{m}$ leads to the motion of the massive particles is
non-geodesic, and an extra force, orthogonal to the four-velocity,
arises \cite{Nojiri5, Papa}. Harko et al \cite{Nojiri5} proposed
an extension of standard general relativity, where the
gravitational Lagrangian is given by an arbitrary function of the
Ricci scalar $R$ and of the trace of the stress-energy tensor $T$
and dubbed this as $f(R,T)$ gravity. The $f(R,T)$ gravity model
depends on a source term, representing the variation of the matter
stress-energy tensor with respect to the metric. A general
expression for this source term is obtained as a function of the
matter Lagrangian $\mathcal{L}_{m}$ \cite{Nojiri5}. In a recent
paper, Myrzakulov \cite{Myrza1} derived exact solutions for a
specific model $f(R,T)=\mu R+\nu T$ and showed that for some
values of the parameters the expansion of our universe can be
accelerated without introducing any dark component. The present
work aims to reconstruct the Ricci dark energy (RDE) \cite{Gao,
Kim, Feng, Fu, Xu} under $f(R,T)$ gravity. Rest of the work is
organized as follows: In section II we have briefly reviewed RDE.
In section III we have presented an overview of $f(R,T)$ gravity.
In section IV we have reconstructed interacting RDE in $f(R,T)$
gravity. We have concluded in section V.

\section{A brief overview of Ricci dark energy}
Gao et.al \cite{Gao} proposed a holographic dark energy model in
which the future event horizon is replaced by the inverse of the
Ricci scalar curvature, and dubbed this model the ``Ricci dark
energy model"(RDE). This model (i) avoids the causality problem
(ii) is phenomenologically viable, and (iii) can solve the
coincidence problem of dark energy \cite{Feng}. The Ricci
curvature of FRW universe is given by \cite{Feng}
\begin{equation}
R=-6\left(\dot{H}+2H^{2}+\frac{k}{a^{2}}\right)
\end{equation}
where, $k$ is the curvature of the universe and $a$ is the scale
factor. The energy density of RDE is given by \cite{Feng1}
\begin{equation}
\rho_{X}=3c^{2}\left(\dot{H}+2H^{2}+\frac{k}{a^{2}}\right)
\end{equation}
 In flat FRW universe, $k=0$ and hence we have
\begin{equation}
\rho_{X}=3c^{2}\left(\dot{H}+2H^{2}\right)
\end{equation}
In the present work we are considering RDE interacting with
pressureless dark matter with energy density $\rho_{m}$. Various
forms of ``interacting" dark energy models have been constructed
in order to fulfil the observational requirements. Plethora of
literatures are available where the interacting dark energies have
been discussed. Several examples of interacting dark energy are
presented in \cite{Jamil, Wu, Kim, Setare, Wang, Karami}. In a
subsequent section we shall consider the interacting RDE in
$f(R,T)$ gravity. The metric of a spatially flat homogeneous and
isotropic universe in FRW model is given by
\begin{equation}
ds^{2}=dt^{2}-a^{2}(t)\left[dr^{2}+r^{2}(d\theta^{2}+sin^{2}\theta
d\phi^{2})\right]
\end{equation}
where $a(t)$ is the scale factor. The Einstein field equations are
given by
\begin{equation}
H^{2}=\frac{1}{3}\rho
\end{equation}
and
\begin{equation}
\dot{H}=-\frac{1}{2}(\rho+p)
\end{equation}
where $\rho$ and $p$ are energy density and isotropic pressure
respectively (choosing $8\pi G=c=1$). The conservation equation is
given by
\begin{equation}
\dot{\rho}+3H(\rho+p)=0
\end{equation}
As we are considering interaction between RDE and dark matter,
\begin{equation}
\rho=\rho_{X}+\rho_{m},~~p=p_{X}
\end{equation}
It should be stated that we are considering pressureless dark
matter, $p_{m}=0$. Since
 the components do not satisfy the
conservation equation separately in the case of interaction, we
need to reconstruct the conservation equation by introducing an
interaction term $Q$. The interaction term could be in any of the
forms \cite{sheykhi}: $Q\propto H\rho_{X}$, $Q\propto H\rho_{m}$
and $Q\propto H(\rho_{X}+\rho_{m})$. In the present paper we take
the interaction term in the second of the three forms mentioned
above. Accordingly the conservation equation is reconstructed as
\begin{equation}
\dot{\rho}_{X}+3H(\rho_{X}+p_{X})=3H\delta\rho_{m}
\end{equation}
\begin{equation}
\dot{\rho}_{m}+3H \rho_{m}=-3H\delta\rho_{m}
\end{equation}
\section{The $f(R,T)$ model}
One of interesting models of $f(R,T)$ gravity is the so-called
$M_{37}$-model. Its action is \cite{Myrza1}
\begin{equation}
S=\int f(R,T)\sqrt{-g}d^{4}x+\int \mathcal{L}_{m}\sqrt{-g}d^{4}x
\end{equation}
where $\mathcal{L}_{m}$ is the matter Lagrangian and $f(R,T)$ is
an arbitrary function of $R$ and $T$, where $R$ is the scalar
curvature and $T$ is the torsion scalar. Here,
\begin{equation}
R=u+6(\dot{H}+2H^{2})
\end{equation}
\begin{equation}
T=v-6H^{2}
\end{equation}
Ref \cite{Myrza1} considered the following model of $f(R,T)$
\begin{equation}
f(R,T)=\mu R+\nu T
\end{equation}
 where $\mu$ and $\nu$ are real constants and $u$ and $v$ are taken as $u=\alpha a^{n}$ and $v=\beta a^{m}$
 with $m, ~n,~\alpha$ and $\beta$ as real constants.
 The equations system
of this $f(R,T)$ is
\begin{equation}
\mu D_{1}+\nu E_{1}+K(\nu T+\mu R)=-2a^{3}\rho
\end{equation}
\begin{equation}
\mu A_{1}+\nu B_{1}+M(\nu T+\mu R)=6a^{2}p
\end{equation}
\begin{equation}
\dot{\rho}+3H (\rho+p)=0
\end{equation}
where,
\begin{equation}
D_{1}=a^{3}\left(6\frac{\ddot{a}}{a}+\dot{a}u_{\dot{\alpha}}\right)
\end{equation}
\begin{equation}
E_{1}=a^{3}\left(-12\frac{\dot{a}^{2}}{a^{2}}+\dot{a}v_{\dot{\alpha}}\right)
\end{equation}
\begin{equation}
K=-a^{3}
\end{equation}
\begin{equation}
A_{1}=12\dot{a}^{2}+6a\ddot{a}+3a^{2}\dot{a}u_{\dot{\alpha}}+a^{3}\dot{u}_{\dot{\alpha}}-a^{3}u_{\alpha}
\end{equation}
\begin{equation}
B_{1}=-24\dot{a}^{2}-12a\ddot{a}+3a^{2}\dot{a}v_{\dot{\alpha}}+a^{3}\dot{v}_{\dot{\alpha}}-a^{3}v_{\alpha}
\end{equation}
\begin{equation}
M=-3a^{2}
\end{equation}
\begin{equation}
R=u+6(\dot{H}+2H^{2})
\end{equation}
\begin{equation}
T=v-6H^{2}
\end{equation}
Subsequently, the modified field equations are obtained as
\cite{Myrza1}
\begin{equation}
3(\mu+\nu)H^{2}+\frac{1}{2}(\mu\alpha a^{n}+\nu\beta a^{m})=\rho
\end{equation}
\begin{equation}
(\mu+\nu)(2\dot{H}+3H^{2})+\frac{\mu\alpha(n+3)}{6}a^{n}+\frac{\nu\beta(m+3)}{6}a^{m}=-p
\end{equation}
\section{Interacting RDE in the $f(R,T)$ gravity}
As stated earlier, $\rho=\rho_{X}+\rho_{m}$ and $p=p_{X}$ are
taken in the equations (26) and (27). From equation (10) we get
\begin{equation}
\rho_{m}=\rho_{m0}a^{-3(1+\delta)}
\end{equation}
Using equations (3) and (28) in the right hand side of the
equation (26) we get the Hubble's parameter as a function of the
scale factor $a$ as
\begin{equation}
H^{2}=C_{1}a^{\frac{2(-2c^{2}+\mu+\nu)}{c^{2}}}+\frac{1}{3a^{3}}\left[\frac{\alpha\mu
a^{3+m}}{c^{2}(4+m)-2(\mu+\nu)}+\frac{\beta\nu
a^{3+n}}{c^{2}(4+n)-2(\mu+\nu)}+\frac{2a^{-3\delta}\rho_{m0}}{c^{2}(-1+3\delta)+2(\mu+\nu)}\right]
\end{equation}
Subsequently we get $\dot{H}$ and $\ddot{H}$ as functions of $a$
as follows
\begin{equation}
\begin{array}{c}
  \dot{H}=\frac{a^{\frac{2(-2c^{2}+\mu+\nu)}{c^{2}}}C_{1}(-2c^{2}+\mu+\nu)}{c^{2}} \\
  +\frac{a^{m}m\alpha\mu}{6(c^{2}(4+m)-2(\mu+\nu))}+\frac{a^{n}n\beta\nu}{6(c^{2}(4+n)-2(\mu+\nu))}-\frac{a^{-3(1+\delta)(1+\delta)\rho_{m0}}}{c^{2}(-1+3\delta)+2(\mu+\nu)} \\
\end{array}
\end{equation}
\begin{equation}
\begin{array}{c}
  \ddot{H}=\frac{H}{6a^{4}}\left[\frac{12a^{\frac{2(\mu+\nu)}{c^{2}}}C_{1}(-2c^{2}+\mu+\nu)^{2}}{c^{4}}\right. \\
  \left.+\frac{a^{4+m}m^{2}\alpha\mu}{c^{2}(4+m)-2(\mu+\delta)}+\frac{a^{4+n}n^{2}\beta\nu}{c^{2}(4+n)-2(\mu+\nu)}+\frac{18a^{1-3\delta}(1+\delta)^{2}\rho_{m0}}{c^{2}(-1+3\delta)+2(\mu+\nu)}\right] \\
\end{array}
\end{equation}
Using (29) and (30) in equation (3) we get the energy density for
RDE under interaction with pressureless dark matter under $f(R,T)$
gravity as
\begin{equation}
\rho_{X}=\frac{1}{2}\left[6a^{\frac{2(-2c^{2}+\mu+\nu)}{c^{2}}}C_{1}(\mu+\nu)+\frac{a^{m}c^{2}\alpha\mu(4+m)}{c^{2}(4+m)-2(\mu+\nu)}+\frac{a^{n}c^{2}\beta\nu(4+n)}{c^{2}(4+n)-2(\mu+\nu)}-\frac{2a^{-3(1+\delta)}c^{2}(-1+3\delta)\rho_{m0}}{c^{2}(-1+3\delta)+2(\mu+\nu)}\right]
\end{equation}
Using the above form of $\rho_{X}$ in the conservation equation
(9) we get the pressure for RDE in the present case as
\begin{equation}
\begin{array}{c}
  p_{X}=\frac{1}{6}\left[-a^{n}(3+n)\alpha \mu-a^{m}(3+m)\beta \nu-6(\mu+\nu)\left\{-\frac{a^{\frac{2(-2c^{2}+\mu+\nu)}{c^{2}}}C_{1}(c^{2}-2(\mu+\nu))}{c^{2}}+\right.\right. \\
  \left.\left.\frac{a^{m}(3+m)\alpha \mu}{3(c^{2}(4+m)-2(\mu+\nu))}+\frac{a^{n}(3+n)\beta\nu}{3(c^{2}(4+n)-2(\mu+\nu))}-\frac{2a^{-3(1+\delta)}\delta\rho_{m0}}{c^{2}(-1+3\delta)+2(\mu+\nu)}\right\}\right]\\
\end{array}
\end{equation}
Using the expressions for energy density and pressure derived
above we get the equation of state parameters
\begin{equation}
w_{X}=\frac{p_{X}}{\rho_{X}}
\end{equation}
and
\begin{equation}
w_{total}=\frac{p_{X}}{\rho_{X}+\rho_{m}};~~(p_{m}=0)
\end{equation}
\begin{figure}[h]
\begin{minipage}{20pc}
\includegraphics[width=20pc]{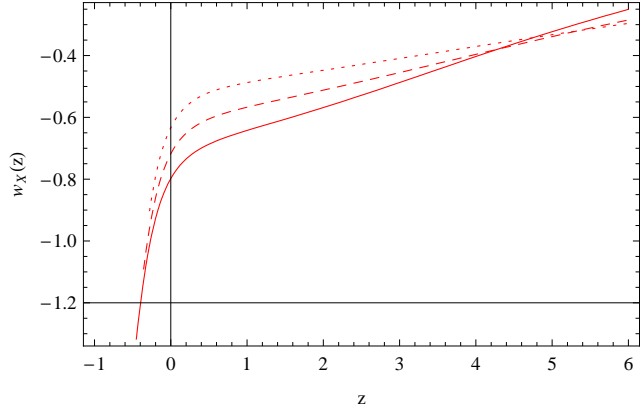}
\caption{\label{label}Behavior of $w_{X}=\frac{p_{X}}{\rho_{X}}$
against redshift $z=a^{-1}-1$.}
\end{minipage}\hspace{2pc}%
\end{figure}

\begin{figure}[h]
\begin{minipage}{20pc}
\includegraphics[width=20pc]{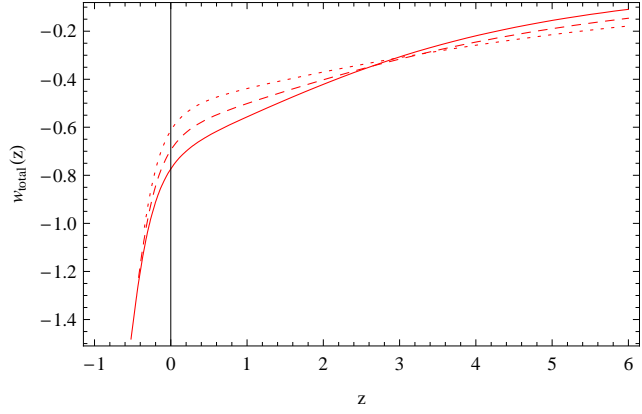}
\caption{\label{label}Behavior of
$w_{total}=\frac{p_{X}}{\rho_{X}+\rho_{m}}$.}
\end{minipage}\hspace{3pc}%
\end{figure}
The deceleration parameter $q$ \cite {Gong} comes out to be
\begin{equation}
q=-\frac{a\ddot{a}}{\dot{a}^{2}}=-1-\frac{\dot{H}}{H^{2}}=-1-\frac{\zeta_{1}}{\zeta_{2}}
\end{equation}
where,
\begin{equation}
\zeta_{1}=\frac{6a^{2(-2c^{2}+\mu+\nu)}C_{1}(-2c^{2}+\mu+\nu)}{c^{2}}+\frac{a^{m}m\alpha
\mu}{c^{2}(4+m)-2(\mu+\nu)}+\frac{a^{n}n\beta
\nu}{c^{2}(4+n)-2(\mu+\nu)}-\frac{6a^{-3(1+\delta)}(1+\delta)\rho_{m0}}{c^{2}(-1+3\delta)+2(\mu+\nu)}
\end{equation}
\begin{equation}
\zeta_{2}=2\left(3a^{\frac{2(-2c^{2}+\mu+\nu)}{c^{2}}}C_{1}+\frac{a^{m}m\alpha\mu}{c^{2}(4+m)-2(\mu+\nu)}+\frac{a^{n}\beta\nu}{c^{2}(4+n)-2(\mu+\nu)}+\frac{2a^{-3(1+\delta)}\rho_{m0}}{c^{2}(-1+3\delta)+2(\mu+\nu)}\right)
\end{equation}
Next, we consider the statefinder parameters $\{r,s\}$
\cite{Sahni} for the present case. Using equations (29), (30) and
(31) we get the statefinder parameters as
\begin{equation}
r =
1+3\frac{\dot{H}}{H^{2}}+\frac{\ddot{H}}{H^{3}}=1+\frac{\varrho_{1}}{\varrho_{2}}
\end{equation}
\begin{equation}
\begin{array}{c}
  \varrho_{1}=\frac{6a^{\frac{2(-2c^{2}+\mu+\nu)}{c^{2}}}C_{1}\{2c^{4}-5c^{2}(\mu+\nu)+2(\mu+\nu)^{2}\}}{c^{4}} \\
  +\frac{a^{m}m(3+m)\alpha\mu}{(c^{2}(4+m)-2(\mu+\nu))}+\frac{a^{n}n(3+n)\beta\nu}{(c^{2}(4+n)-2(\mu+\nu))}-\frac{18a^{-3(1+\delta)}(1+\delta)\delta\rho_{m0}}{c^{2}(-1+3\delta)+2(\mu+\nu)} \\
\end{array}
\end{equation}
\begin{equation}
\begin{array}{c}
  \varrho_{2}=2\left[3C_{1}a^{\frac{2(-2c^{2}+\mu+\nu)}{c^{2}}}+\frac{1}{a^{3}}\left\{\frac{\alpha\mu
a^{3+m}}{c^{2}(4+m)-2(\mu+\nu)}+\right.\right. \\
  \left.\left.\frac{\beta\nu
a^{3+n}}{c^{2}(4+n)-2(\mu+\nu)}+\frac{2a^{-3\delta}\rho_{m0}}{c^{2}(-1+3\delta)+2(\mu+\nu)}\right\}\right] \\
\end{array}
\end{equation}
\begin{equation}
s=-\frac{3H\dot{H}+\ddot{H}}{3H(2\dot{H}+3H^{2})}=\frac{\xi_{1}}{\xi_{2}}
\end{equation}
where,
\begin{equation}
\begin{array}{c}
  \xi_{1}=-\left(\frac{a^{m}m(3+m)\alpha\mu}{6(c^{2}(4+m)-2(\mu+\nu))}+\frac{a^{n}n(3+n)\beta\nu}{6(c^{2}(4+n)-2(\mu+\nu))}\right. \\
 \left.
 +\frac{1}{c^{4}}a^{\frac{2(-2c^{2}+\mu+\nu)}{c^{2}}}C_{1}(2c^{4}-5c^{2}(\mu+\nu)+2(\mu+\nu)^{2})+\frac{3a^{-3(1+\delta)}\delta(1+\delta)\rho_{m0}}{c^{2}(-1+3\delta)+2(\mu+\nu)}\right)
 \\
\end{array}
\end{equation}
\begin{equation}
\begin{array}{c}
  \xi_{2}=3\left(-\frac{1}{c^{2}}a^{\frac{2(-2c^{2}+\mu+\nu)}{c^{2}}}+\frac{a^{m}(3+m)\alpha\mu}{3(c^{2}(4+m)-2(\mu+\nu))}\right.\\
   \left.+\frac{a^{n}(3+n)\beta\nu}{3(c^{2}(4+n)-2(\mu+\nu))}-\frac{2a^{-3(1+\delta)}\delta\rho_{m0}}{c^{2}(-1+3\delta)+2(\mu+\nu)}\right)\\
\end{array}
\end{equation}
\section{Discussions}
In figure 1 we have presented the EoS parameter
$w_{X}=\frac{p_{X}}{\rho_{X}}$ for RDE under $f(R,T)$ gravity
against redshift $z=a^{-1}-1$. In this and the subsequent figures
the solid, dashed and dotted lines would correspond to
$c^{2}<,~=,~>0.5$ respectively. Figure 1 shows that for all values
of $c^{2}$ the EoS parameter transits from $w_{X}>-1$ to
$w_{X}<-1$ i.e. from quintessence to phantom. From this figure we
see that at early times, roughly $z>2$, the EoS approaches 0;
i.e., in this model the dark energy behaves like dust matter
during most of the epoch of matter domination. The EoS crosses
phantom crossing $w_{X}=-1$ at $z\approx-0.2$ and in the distant
future, the equation of state approaches $w_{X}=-1.2$, the
Universe evolves into the phantom-dominated epoch. For this model,
the EoS crosses $-1$ , so it may be classified as a ``quintom".
Thus, the interacting RDE behaves like quintom in the $f(R,T)$
gravity model proposed by \cite{Myrza1}. In figure 2 we have
plotted $p_{total}=\frac{p_{X}}{\rho_{X}+\rho_{m}}$, where we
found similar crossing of the phantom divide $w_{total}=-1$ and
transition from $w_{total}>-1$ at higher redshift to
$w_{total}<-1$ at lower redshifts. It might be stated that we have
chosen the model parameters as
$\alpha=1.2,~\beta=1.2,~\nu=0.3,~\mu=0.4,~n=3,C_{1}=3.02,~\delta=0.05,~
\rho_{m0}=0.23$ and $m=2$. In all the figures, the solid, dashed
and dotted lines correspond to $c^{2}<0.5,~=0.5$ and $>0.5$
respectively.

\begin{figure}[h]
\begin{minipage}{20pc}
\includegraphics[width=20pc]{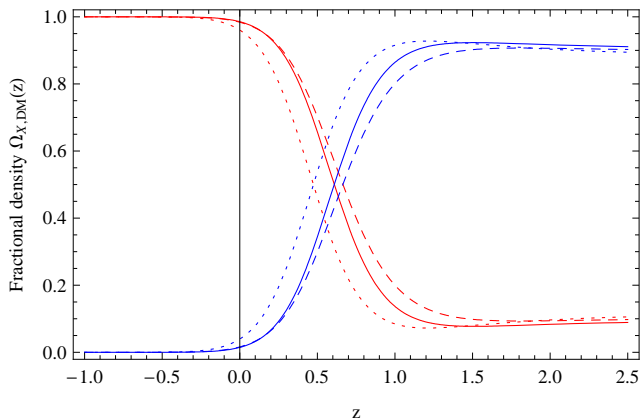}
\caption{\label{label}Behavior of fractional densities
$\Omega_{X}=\frac{\rho_{X}}{3\tilde{H}^{2}}$ (red lines) and
$\Omega_{m}=\frac{\rho_{m}}{3\tilde{H}^{2}}$ (blue lines) with
evolution of the universe.}
\end{minipage}\hspace{2pc}%
\end{figure}

In figure 3 we have plotted fractional densities
$\Omega_{X}=\frac{\rho_{X}}{3\tilde{H}^{2}}$ and
$\Omega_{m}=\frac{\rho_{m}}{3\tilde{H}^{2}}$ against redshift $z$.
where, $ \tilde{H}^{2}=(\mu+\nu)H^{2}+\frac{1}{6}(\mu\alpha
(1+z)^{-n}+\nu\beta (1+z)^{-m})$. We observe that at from higher
to lower redshifts the fractional density $\Omega_{X}$ of RDE is
increasing, while the fractional density of dark matter is
decreasing. This indicates the transition from the matter
dominated to dark energy dominated universe. At very early stage
of universe $z>2$, the fractional density of dark energy
$\Omega_{X}$ is dominated by fractional density of dark matter
$\Omega_{DM}$. After $z=2$, the $\Omega_{X}$ starts showing an
increasing pattern and $\Omega_{DM}$ starts showing a decaying
pattern. This indicates the gradual transition from matter
dominated era to the dark energy dominated era. We denote the
cross-over point of the fractional densities by $z_{cross}$, where
$\Omega_{X}=\Omega_{DM}$. For $c^{2}<0.5,~=0.5$ and $>0.5$ the
$z_{cross}\approx0.6,~0.75$ and $0.5$ respectively. It is also
observed that in the early universe the density contribution of
dark energy can occupy roughly 20$\%$-30$\%$. However, at this
stage the dark energy behaves like dust matter. So, effectively
speaking, the matter density contribution is still 100$\%$.
Finally, from figure 3 our observation is that RDE in $f(R,T)$
gravity is capable of achieving the transition from
matter-dominated to dark energy-dominated universe.\\
\begin{figure}[h]
\begin{minipage}{20pc}
\includegraphics[width=20pc]{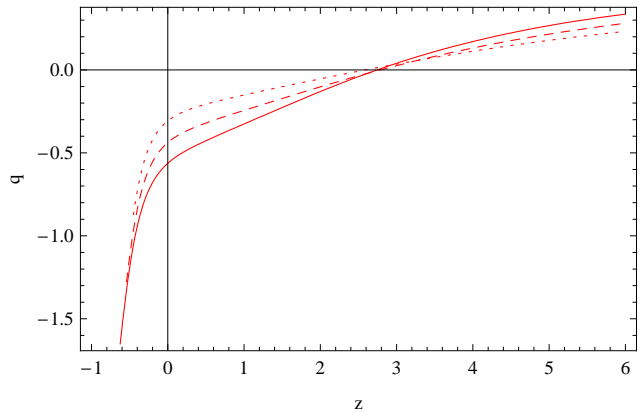}
\caption{\label{label}Behavior of deceleration parameter $q$ (Eq.
36).}
\end{minipage}\hspace{3pc}%
\end{figure}
In figure 4 we have plotted the deceleration parameter $q$ as a
function of the redshift $z$. We observe that at very early stage,
roughly $z>2$, $q>0$ i.e. the decelerated universe. At
$z\approx2.5$ the deceleration parameter transits from positive to
negative level. That is, the universe gradually transits from
decelerated to accelerated stage. At later stage $q=-1.5$. Thus,
we observe that it is possible to achieve the accelerated phase of
the universe from decelerated phase for RDE under $f(R,T)$
gravity.\\
\begin{figure}[h]
\begin{minipage}{20pc}
\includegraphics[width=20pc]{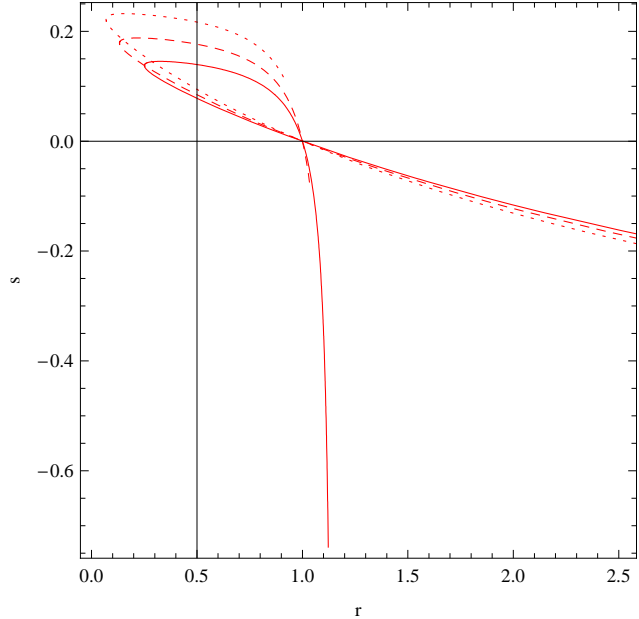}
\caption{\label{label}The $r-s$ trajectory (Eqs. 39 and 42).}
\end{minipage}\hspace{3pc}%
\end{figure}
In figure 5 we have created the $r-s$ trajectories for the three
values of $c^{2}$ under consideration. Sahni et al. \cite{Sahni}
demonstrated that the statefinder diagnostic could effectively
discriminate different forms of dark energy. Cosmological
constant, quintessence, Chaplygin gas, and braneworld models were
investigated by \cite{Alam} using the statefinder diagnostics and
it was observed that the statefinder pair could differentiate
between these models. An investigation on statefinder parameters
for differentiating between dark energy and modified gravity was
carried out in \cite{Wang1}. Statefinder diagnostics for $f(T)$
gravity has been investigated in \cite{WuYu}. In the $\{r,s\}$
plane, $s>0$ corresponds to a quintessence like dark energy and
$s<0$ corresponds to a phantom-like dark energy, and an evolution
from phantom to quintessence or inverse is given by a crossing of
the point $(r=1,s=0)$ in $\{r,s\}$ plane \cite{WuYu}. Also, the
fixed point $(r=1,s=0)$ corresponds to $\Lambda$CDM scenario. In
figure 5 we clearly observe a transition from quintessence to
phantom as the $r-s$ trajectory transits from positive to negative
sides of $s$ after crossing the $(r=1,s=0)$ point. Also, we find
that, for finite $r$, $s\rightarrow-\infty$ that corresponds to
dust phase. Thus, the interacting RDE in $f(,T)$ gravity with
$f(R,T)=\mu R+\nu T$ interpolates between dust and $\Lambda$CDM
phases of the universe. Also, the statefinder diagnostics supports
the quintom-like behavior of the equation of state.

\section{Concluding remarks}
In this work we considered interacting Ricci dark energy in
$f(R,T)=\mu R+\nu T$ gravity. We have observed that the EoS
parameter exhibits quintom like behavior for this model. Also, the
transition from matter dominated to dark energy dominated universe
is achievable by this model. The deceleration parameter have
exhibited a transition from positive to negative sign, thereby
showing the evolution of the universe from deceleration to
acceleration. The statefinder diagnostics have been investigated
and an interpolation between dust and $\Lambda$CDM phases of the
universe has been observed under this model.

\end{document}